\documentclass[a4paper,12pt]{article}
\usepackage{tipa}
\usepackage{amsmath}
\usepackage{amssymb}
\usepackage{amsmath,amsthm,amssymb,amscd}
\usepackage{latexsym}
\usepackage{indentfirst}
\usepackage{subfigure}
\usepackage{graphicx}
\usepackage[top=1in,bottom=1in,left=1.25in,right=1.25in]{geometry}
\makeatother
\date{}
\theoremstyle{plain}

\begin{document}
\bibliographystyle{plain}

\title{
  \bf Domain wall of the totally asymmetric exclusion process without particle number conservation  
}

\author{{Yunxin Zhang}\thanks{
School of Mathematical Sciences, Fudan University,  Shanghai 200433,
China  (E-Mail: xyz@fudan.edu.cn)}
\thanks{Shanghai Key Laboratory for Contemporary Applied Mathematics, Fudan University},\
\thanks{Centre for Computational Systems Biology, Fudan University}
}


\maketitle \baselineskip=6mm

\begin{abstract}
In this research, the totally asymmetric exclusion process without
particle number conservation is discussed. Based on the mean field
approximation and the Rankine-Hugoniot condition, the necessary and
sufficient conditions of the existence of the domain wall have been
obtained. Moreover, the properties of the domain wall, including the
location and height, have been studied theoretically. All the
theoretical results are demonstrated by the computer simulations.

\vspace{2em} \noindent \textit{PACS}: 87.16.Nn, 87.16.A-, 05.60.-k,
05.70.Ln

\vspace{2em} \noindent \textit{Keywords}: TASEP; domain wall;
molecular motors
\end{abstract}
%

\section{Introduction}

One-dimensional driven diffusion system is a very interesting
research topic in recent years. They were shown to exhibit boundary
induced phase transitions \cite{krag}, and phase separation
\cite{Evans, Parmeggiani, Derrida1998}. In Ref. \cite{Mirin}, the
effect of a single detachment site in the bulk of an asymmetric
simple exclusion process (ASEP) was studied. In Refs
\cite{Parmeggiani, Willmann}, the interplay of the simplest
one-dimensional driven model, the totally asymmetric exclusion
process (TASEP) with local absorption/desorption kinetics of single
particles acting at all sites, termed \lq\lq Langmuir kinetics" (LK)
was considered. These models were inspired by the dynamics of motor
proteins \cite{Alberts} \cite{Astumian1997} \cite{Howard2001}
\cite{Zhang2008} \cite{David2007}, which move along cytoskeletal
filaments in a certain preferred direction while detachment and
attachment can also occur between the cytoplasm and the filament,
and, in a very different setting, by dynamics of limit orders in a
stock exchange market. Being an equilibrium process, LK is well
understood, while the combined process of TASEP and LK showed the
new feature of a localized domain wall in the density profile of the
stationary state \cite{Parmeggiani}.

The TASEP is defined on a one-dimensional lattice of size $N$. Each
site can either be empty or occupied by one particle. In the bulk,
particles can hop from site $i$ to site $i+1$ with unit rate,
provided the target site is empty. At site 1, particles can enter
the lattice from a reservoir with density $\alpha$ , provided the
site is empty. They can leave the system from site $N$ into a
reservoir of density $\beta$ with rate $1-\beta$ . Thus in the
interior of the lattice, the particle number is a conserved
quantity. The phase diagram and steady states of the TASEP as a
function of the boundary rates are known exactly \cite{Liggett}
\cite{Schutz} \cite{Derrida}. Furthermore, a theory of boundary
induced phase transitions exists, which explains the phase diagram
quantitatively in terms of the dynamics of domain wall
\cite{Kolomeisky}. In the field of TASEP, Joel L. Lebowitz {\sl et
al} have done many excellent research and obtained much useful
theoretical results \cite{Derrida1997, Derrida1993, Ayyer2008}.

Similar as in \cite{Willmann}, in this research we equip the system
with the additional feature of local particle creation at empty
sites with rate $\omega_a$ and annihilation with rate $\omega_d$. In
the thermodynamic limit $N\to\infty$, the case of the local rates
being of the order of $1/N$ is the most interesting one
\cite{Popkov}. It turns out that the presence of the kinetic rates
significantly change the picture of TASEP, producing a completely
reorganized phase diagram. In \cite{Parmeggiani}, the authors showed
by computer simulations and mean-field arguments that, in this
nonconserved dynamics, one can have phase coexistence where low and
high density phases are separated by stable discontinuities (domain
wall) in the density profile. Recently, this dynamics was also
studied theoretically in  \cite{Sutapa2007} \cite{Sutapa2005}.

Up to now, the properties of the totally asymmetric exclusion
process without particle number conservation have not been well
studied theoretically, though many of which have been found in the
computer simulations. It is no doubt that some particular properties
of this process are very difficult to be found only by computer
simulations, and the right theoretical analysis should be given to
the full understanding of which. In this research, the necessary and
sufficient conditions of the existence of domain wall and the
properties of the domain wall, for this totally asymmetric exclusion
process without particle number conservation, will be theoretically
discussed. Two basic questions will be answered theoretically: (1)
when and where does the domain wall exist? (2) How do the location
and height of the domain wall change as the parameters change?

This paper is organized as following. In the next section, the
mathematical model and some basic results of this process are
introduced. The necessary and sufficient conditions of the existence
of the domain wall will be given in section 3, then the properties
of the domain wall, including the location and the height, will be
discussed theoretically in section 4. Finally, some concluding
remarks are given in the last section.

\section{Mean field approximation}

The equations of the bulk dynamics of the totally asymmetric
exclusive process with particle creation and annihilation are the
following \cite{Evans, Parmeggiani}
\begin{equation}\label{Eq1}
\frac{dn_i}{dt}=n_{i-1}(1-n_i)-n_i(1-n_{i+1})+\omega_a
(1-n_i)-\omega_d n_i\qquad 2\le i\le N-1
\end{equation}
while at the boundaries
\begin{equation}
\begin{aligned}
&\frac{dn_1}{dt}=\alpha(1-n_1)-n_1(1-n_2)\cr
&\frac{dn_N}{dt}=n_{N-1}(1-n_N)-\beta n_N
\end{aligned}
\end{equation}
where the occupation numbers $n_i=1$ for a site occupied by a
particle and $n_i=0$ for an empty site. For fixed total length $L=1$
and $N\to\infty$ one gets the differential equation for the average
profile in the stationary state \cite{Parmeggiani} \cite{Kolomeisky}
\begin{equation}\label{Eq2}
(2\rho(x)-1)\rho'(x)=(\Omega_a+\Omega_d)\rho(x)-\Omega_a\quad 0< x<
1
\end{equation}
where the reduced rates $\Omega_a=N\omega_a, \Omega_d=N\omega_d $
(because of the particle-hole symmetry, we restrict the discussion
to the case $\Omega_a\ge\Omega_d$ \cite{Parmeggiani}). In the
following, let $\rho_{l\alpha}$ and $\rho_{r\beta}$ (or $\rho_l$ and
$\rho_r$ for simplicity) be the solutions of Eq. (\ref{Eq2}) with
boundary conditions $\rho_{l\alpha}(0)=\alpha$ and
$\rho_{r\beta}(1)=1-\beta$ respectively. It is to say,
$\rho_{l\alpha}$ and $\rho_{r\beta}$ satisfy the following two
equations respectively (see \cite{Evans1})
\begin{equation}\label{slol}
\frac{2(\rho_{l\alpha}-\alpha)}{(K+1)\Omega_d}+\frac{K-1}{(K+1)^2\Omega_d}\ln\left|\frac{K-(K+1)\rho_{l\alpha}}{K-(K+1)\alpha}\right|=x
\end{equation}
\begin{equation}\label{slor}
\frac{2(1-\beta-\rho_{r\beta})}{(K+1)\Omega_d}+\frac{K-1}{(K+1)^2\Omega_d}\ln\left|\frac{K-(K+1)(1-\beta)}{K-(K+1)\rho_{r\beta}}\right|=1-x
\end{equation}
where $K:=\frac{\Omega_a}{\Omega_d}\ge 1$.

\section {The existence of the domain wall}
\begin{figure}
  \includegraphics[width=200pt]{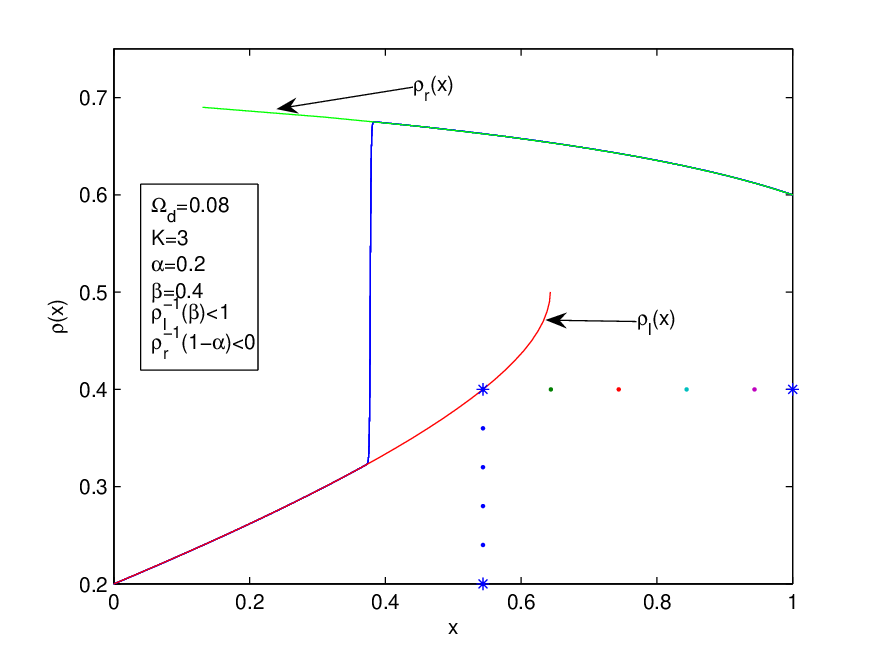}\includegraphics[width=200pt]{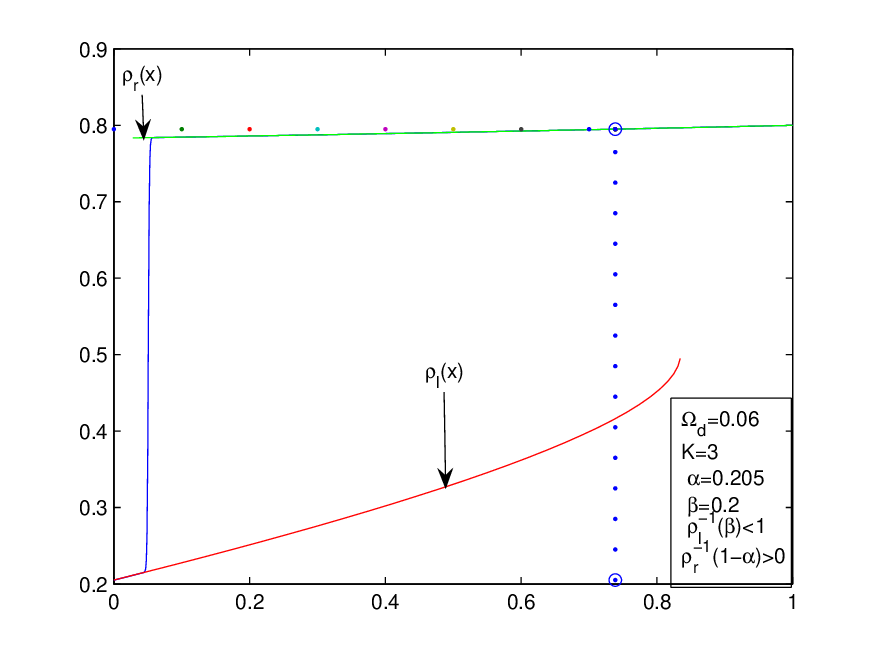}
  \caption{For $0\le \alpha\le 0.5$ and $0.5\le 1-\beta\le\frac{K}{K+1}$: if $\rho_{l\alpha}^{-1}(\beta)<1, \rho_{r\beta}^{-1}(1-\alpha)<0$, there exists domain wall
  in $(0, 1)${\bf (left)}. For $0\le \alpha\le 0.5, \frac{K}{K+1}\le 1-\beta\le 1$: if $\rho_{l\alpha}^{-1}(\beta)\le 1\quad \text{and}\quad
\rho_{r\beta}^{-1}(1-\alpha)\ge 0$, there exists domain wall in $(0,
1)$
  {\bf (right)}.}\label{Figure 1}
\end{figure}

Due to the Rankine-Hugoniot condition \cite{Courant}, at the
location $x_S$ of the domain wall of Eqn. (\ref{Eq2}),
\begin{equation}\label{cond1}
(\rho_+^2-\rho_+)=(\rho_-^2-\rho_-)
\end{equation}
should be satisfied, where $\rho_-=\lim\limits_{x\to x_S-}\rho(x)$,
$\rho_+=\lim\limits_{x\to x_S+}\rho(x)$. It can be easily found that
the condition (\ref{cond1}) can be simplified as
\begin{equation}\label{cond2}
\rho_-+\rho_+=1
\end{equation}
Therefore, to the Eqn. (\ref{Eq2}) with boundary conditions
$\rho(0)=\alpha$ and $\rho(1)=1-\beta$, there exists domain wall in
the interval $(0, 1)$ if and only if there exists a location
$0<x_S<1$ at which $\rho_-+\rho_+=1$, moreover,
\begin{equation}\label{Eq8}
\rho(x)=\left\{
\begin{aligned}
&\rho_l(x)\qquad 0\le x<x_S\cr &\rho_r(x)\qquad x_S<x\le 1
\end{aligned}\right.
\end{equation}
From the conditions (\ref{cond2}) (\ref{Eq8}) and the Eqns.
(\ref{slol}) (\ref{slor}), it is not difficult to obtain the
following theoretical results:

\noindent {\bf(I)} For $0\le\alpha\le 0.5, 0\le
1-\beta\le\frac{K}{K+1}$, the necessary and sufficient conditions of
the existence of domain wall are (see Fig. \ref{Figure 1} {\bf
(left)})
\begin{equation}
\rho^{-1}_{l\alpha}(\gamma)\le 1\quad\text{and}\quad
\rho^{-1}_{r\gamma}(1-\alpha)\le 0\qquad \text{with}\quad
\gamma=\min(0.5,\ \beta)
\end{equation}

\noindent {\bf(II)} For $0\le\alpha\le 0.5, \frac{K}{K+1}\le
1-\beta\le 1$, the necessary and sufficient conditions of the
existence of domain wall are (see Fig. \ref{Figure 1} {\bf (right)})
\begin{equation}
\rho_{l\alpha}^{-1}(\beta)\le 1\quad \text{and}\quad
\rho_{r\beta}^{-1}(1-\alpha)\ge 0
\end{equation}

\noindent {\bf(III)} For $0.5\le\alpha\le 1$, there no domain wall
exists in $(0, 1)$ (for details, see \cite{Zhang}).

\section{The properties of the domain wall  }
In this section, we will discuss the properties of the location
$x_S$ and the height $2\epsilon$ of the domain wall, which can be
regarded as functions of the parameters $K, \Omega_d, \alpha,
\beta$. Where $2\epsilon=|\rho_+-\rho_-|=|2\rho_+-1|$. In the
following, we assume $0\le \alpha\le 0.5$ (which is the necessary
condition of the existence of domain wall), $0\le \beta\le 0.5$ (if
$\beta>0.5$ and the domain wall exists, it is equivalent to the case
in which $\beta=0.5$, see \cite{Zhang}), $\Omega_d\ge 0$ and $K\ge
3$. For the sake of the convenience, we define the following
functions
\begin{equation}
\begin{aligned}
&A=A(\epsilon,K):=2\epsilon+2K\epsilon-K+1\qquad
&B=B(\epsilon,K):=2\epsilon+2K\epsilon+K-1\cr &C=C(K,
\beta):=K\beta+\beta-1 &D=D(K,\alpha):=K\alpha+\alpha-K\qquad\cr
&E=E(\epsilon,\alpha):=1-2\alpha-2\epsilon
&F=F(\epsilon,\beta):=1-2\beta-2\epsilon\quad\qquad\cr
\end{aligned}
\end{equation}

\subsection{The Properties of the Location of the domain wall  }
\begin{figure}
  \includegraphics[width=200pt]{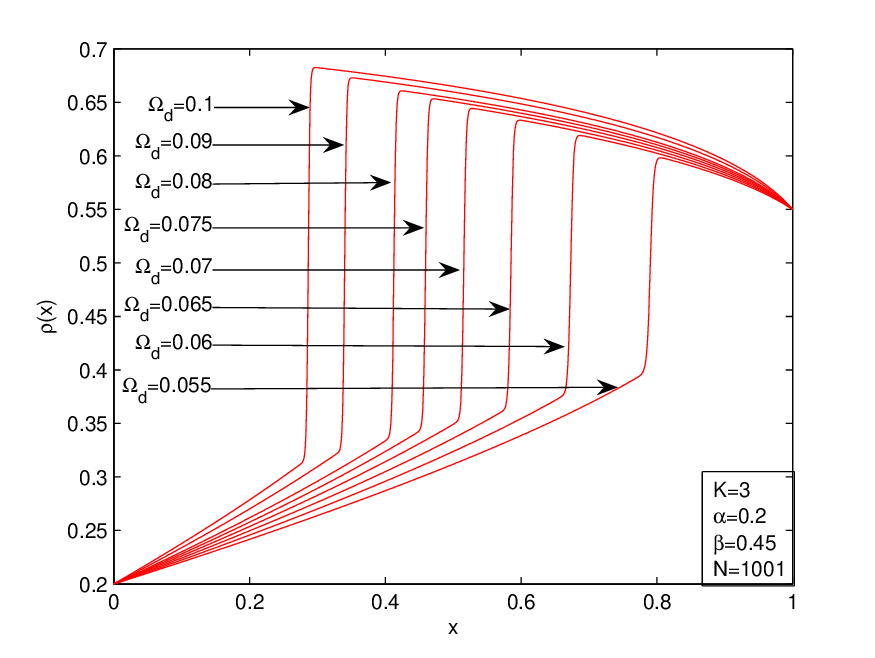}\includegraphics[width=200pt]{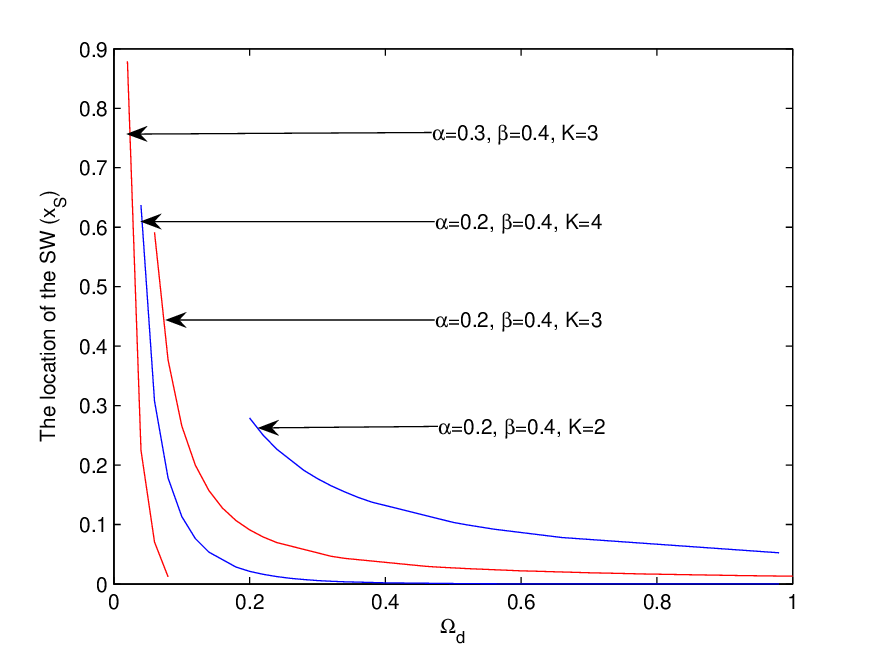}
  \caption{The relationship between the location of the domain wall and the parameter $\Omega_d$, for $0\le \alpha\le 0.5, \frac{1}{K+1}\le \beta\le 0.5$ {\bf
  (left)}. $x_S(\Omega_d)$ is monotonously decreased as a function of $\Omega_d$ for $0\le \alpha\le 0.5, \frac{1}{K+1}\le \beta\le 0.5$  {\bf (right)}.}
\label{Figure 9}
\end{figure}

\begin{figure}
  \includegraphics[width=200pt]{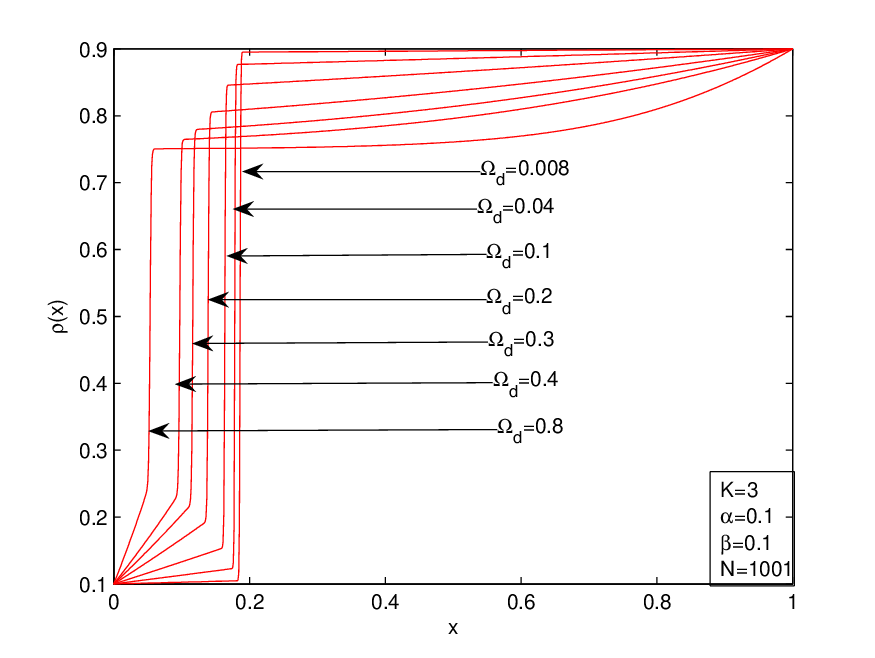}\includegraphics[width=200pt]{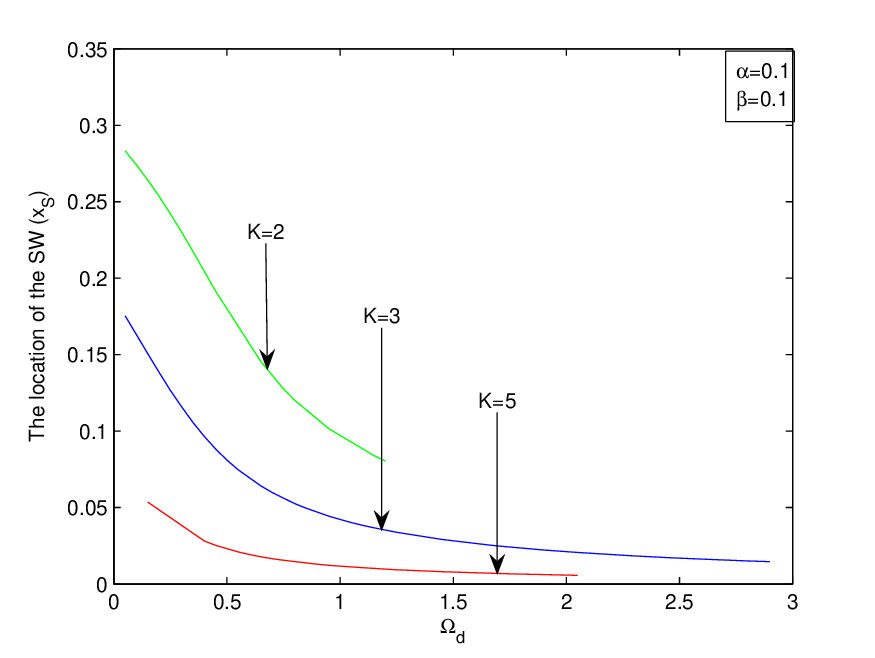}
  \caption{ The relationship between the location of the domain wall and the parameter $\Omega_d$:
$0\le \alpha\le 0.5, 0\le \beta<\frac{1}{K+1}$ {\bf  (left)}.
$x_S(\Omega_d)$ is monotonously decreased as a function of
$\Omega_d$ for $0\le \alpha\le 0.5, 0\le \beta<\frac{1}{K+1}$   {\bf
(right)}. In all the computer simulations, $2\epsilon
(K+1)(1-2x_S)\le (K-1)$ is satisfied.} \label{Figure 10}
\end{figure}

\begin{figure}
  \includegraphics[width=200pt]{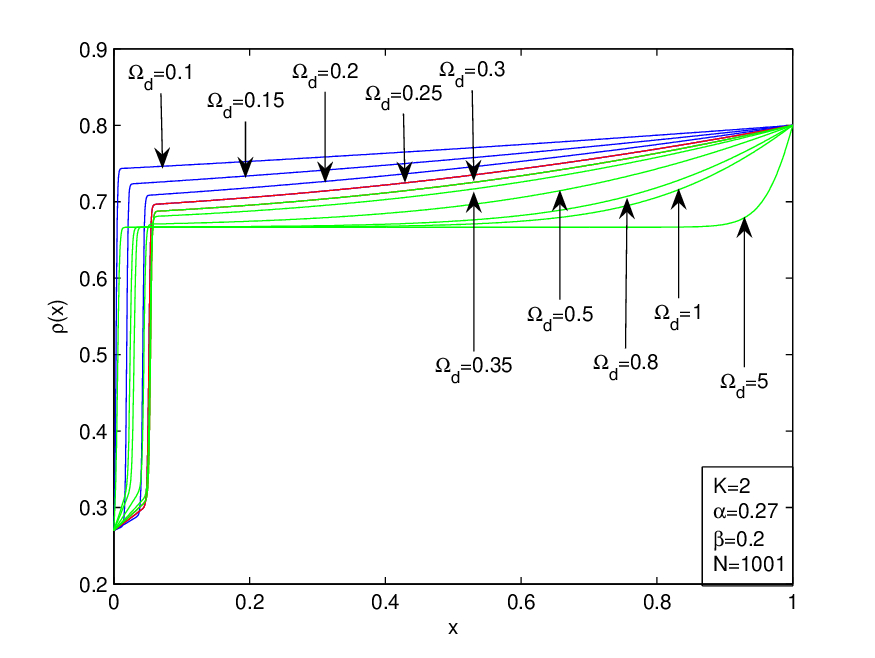}\includegraphics[width=200pt]{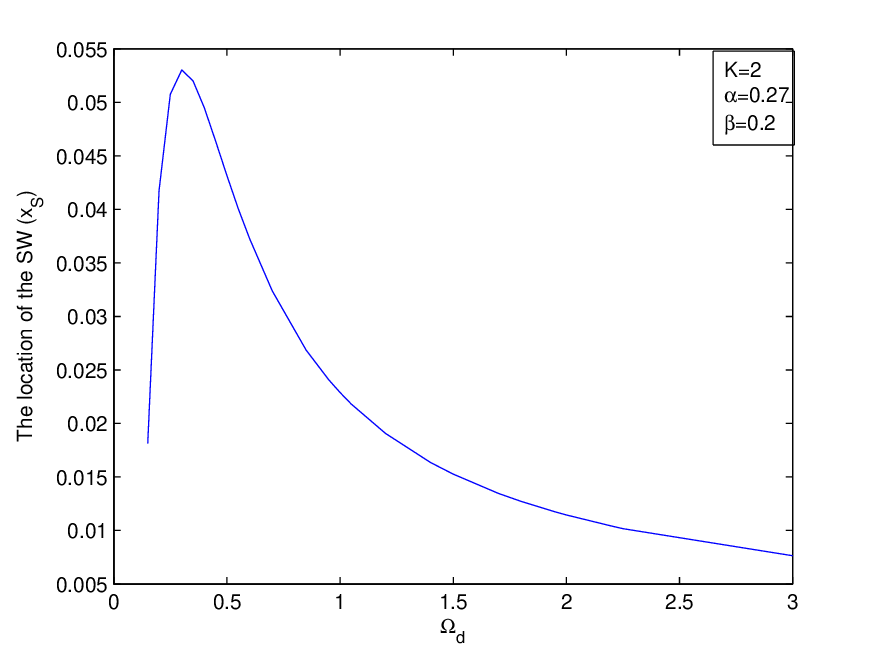}
  \caption{The relationship between the location of the domain wall and the parameter $\Omega_d$ for $0\le \alpha\le 0.5, 0\le \beta<\frac{1}{K+1}$:
 monotonously increased as a function of the parameter $\Omega_d$ when $\Omega_d$ is small, then monotonously decreased when $\Omega_d$ is large enough {\bf (left)}.
The corresponding figure of the function $x_S(\Omega_d)$. {\bf
(right)}. At the critical point, $2\epsilon (K+1)(1-2x_S)=(K-1)$.}
\label{Figure 11}
\end{figure}

From Eqns. (\ref{slol}) and (\ref{slor}), we can get the following
theoretical results:

 \noindent {\bf (a) }
\begin{equation}
\frac{\partial x_S}{\partial \Omega_d}=\frac{A(\epsilon,K)-4\epsilon
x_S(K+1)}{4\epsilon\Omega_d(K+1)}=\frac{2\epsilon
(K+1)(1-2x_S)-(K-1)}{4\epsilon\Omega_d(K+1)}
\end{equation}
For $\frac{1}{K+1}\le \beta\le 0.5$, i.e. $0.5\le 1-\beta\le
\frac{K}{K+1}$, it can be easily proved that $2\epsilon\le
\frac{K-1}{K+1}$, which implies $A(\epsilon,K)\le 0$, so
$\frac{\partial x_S}{\partial \Omega_d}<0$. It is to say that the
location $x_S$ of the domain wall is monotonously decreased as a
function of the parameter $\Omega_d$ (the corresponding computer
simulations are plotted in Figure \ref{Figure 9}).
For $0\le \beta<\frac{1}{K+1}$, i.e. $\frac{K}{K+1}< 1-\beta\le 1$,
\begin{equation}\label{eq27}
\frac{\partial x_S}{\partial \Omega_d}=\frac{2\epsilon
(K+1)(1-2x_S)-(K-1)}{4\epsilon\Omega_d(K+1)}
\end{equation}
if $2\epsilon (K+1)(1-2x_S)\le (K-1)$, the location $x_S$ of the
domain wall is also decreased as the parameter $\Omega_d$ increases
(Figure \ref{Figure 10}).
Otherwise, the location $x_S$ of the domain wall is increased as the
parameter $\Omega_d$ increases (Figure \ref{Figure 11}).

\begin{figure}
  \includegraphics[width=200pt]{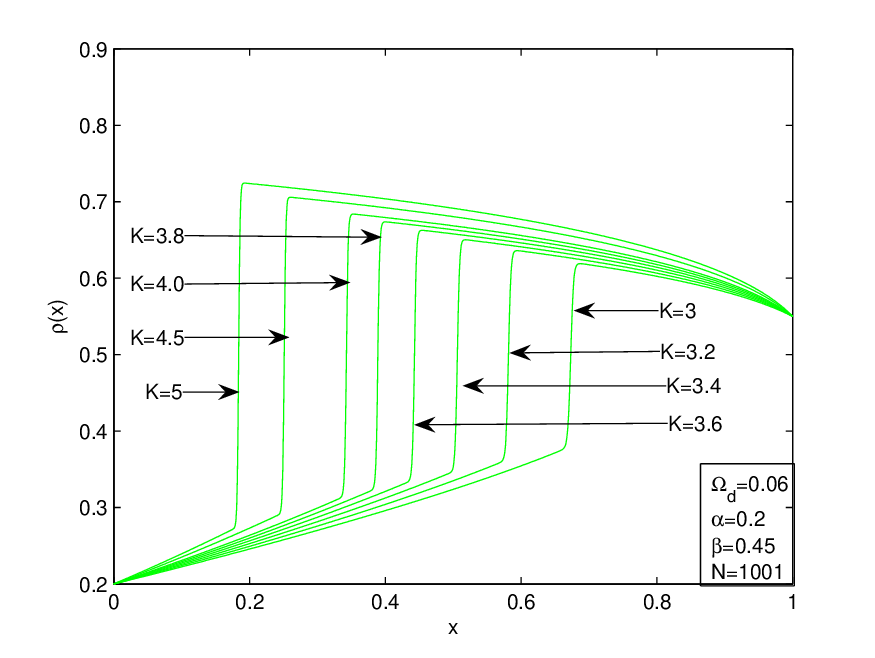}\includegraphics[width=200pt]{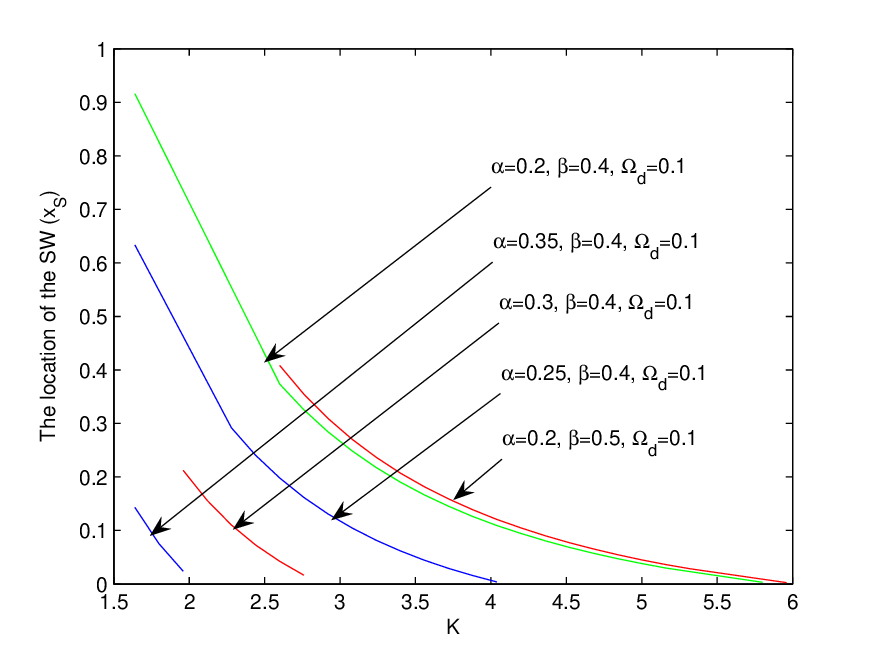}
  \caption{For $0\le \alpha\le 0.5, \frac{1}{K+1}\le\beta\le 0.5$ the location
$x_S$ of the domain wall is decreased as the parameter $K$ increases
{\bf (left)}. The figure of the function $x_S(K)$ for $0\le
\alpha\le 0.5, \frac{1}{K+1}\le\beta\le 0.5$ {\bf (right)}.}
\label{Figure 12}
\end{figure}

\begin{figure}
  \includegraphics[width=200pt]{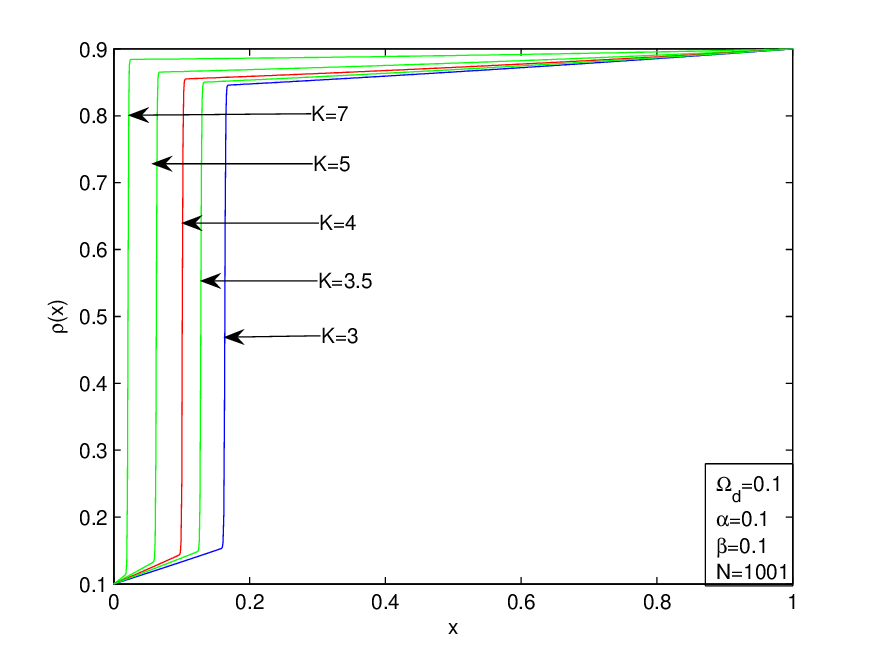}\includegraphics[width=200pt]{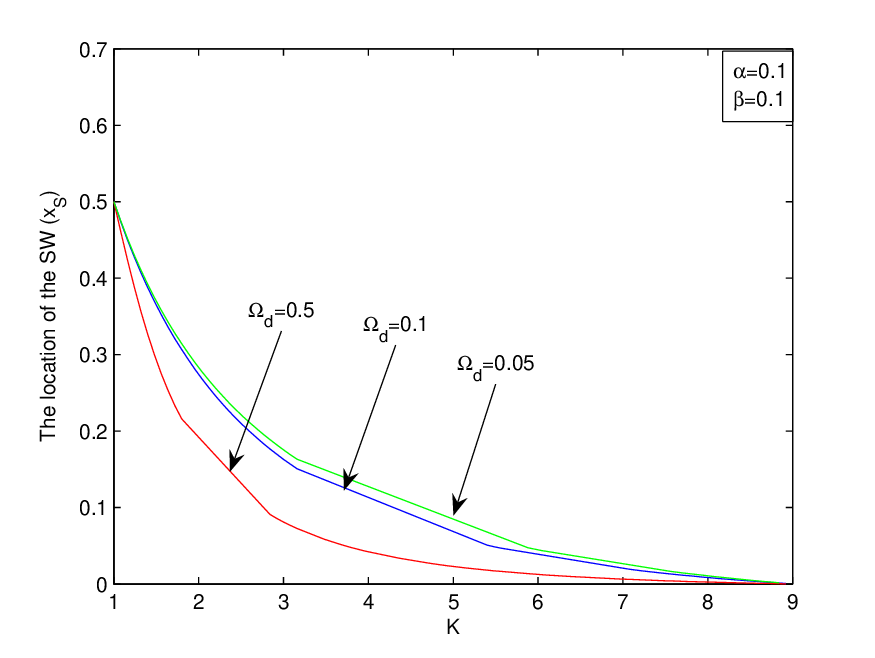}
  \caption{For $0\le \alpha\le 0.5, 0\le
\beta<\frac{1}{K+1}, 2\epsilon (K+1)(1-2x_S)\le (K-1)$ the location
$x_S$ of the domain wall is decreased as the parameter $K$ increase
{\bf (left)}.
 The figure of the function $x_S(K)$ for  $0\le \alpha\le 0.5, 0\le
\beta<\frac{1}{K+1}, 2\epsilon (K+1)(1-2x_S)\le (K-1)$ {\bf
(right)}.} \label{Figure 13}
\end{figure}

\noindent {\bf (b) }
\begin{equation}\label{eq28}
\begin{aligned}
\frac{\partial x_S}{\partial
K}=&-\frac{1}{4(K+1)\epsilon}\left[\frac{K-3}{K^2-1}[(A+B)x_S-A]\right.\cr
&+\left.\frac{[(K^2-1)(2\alpha-1)+4\epsilon(K+1)D]CE-[4\epsilon(K+1)C+2(K+1)(1-2\beta)]DF}{\Omega_d(K+1)^2(K-1)CD}\right]
\end{aligned}
\end{equation}
For $0\le \alpha\le 0.5, \frac{1}{K+1}\le \beta\le 0.5$, it can be
verified that
\begin{equation}\label{eq29}
A\le 0\quad A+B>0\quad C\ge 0\quad D\le 0\quad E\ge 0\quad F\le 0
\end{equation}
which imply
\begin{equation}\label{eq30}
\frac{\partial x_S}{\partial K}\le 0\qquad \text{for}\quad K\ge 3
\end{equation}
i.e. the location $x_S$ of the domain wall is monotonously decreased
as a function of the parameter $K$ for $K\ge 3, 0\le \alpha\le 0.5,
\frac{1}{K+1}\le \beta\le 0.5$ (Figure \ref{Figure 12}).
At the same time, Eq. (\ref{eq28}) can be reformulated as
\begin{equation}\label{eq31}
\begin{aligned}
\frac{\partial x_S}{\partial
K}=&-\frac{1}{4(K+1)\epsilon}\frac{1}{\Omega_d(K+1)^2(K-1)CD}\left\{(K-3)\Omega_d[2\epsilon(K+1)(2x_S-1)+(K-1)]\right.\cr
&+\left.8\epsilon(\beta-\alpha)+(K^2-1)(2\alpha-1)CE+(K-1)(2\beta-1)DF\right\}
\end{aligned}
\end{equation}
For $0\le \beta<\frac{1}{K+1}$, it is easy to verify
\begin{equation}\label{eq32}
C\le 0\quad D\le 0\quad E\ge 0\quad F\ge 0
\end{equation}
if
\begin{equation}\label{eq33}
2\epsilon (K+1)(1-2x_S)\le
(K-1)+\frac{8\epsilon(\beta-\alpha)+(K^2-1)(2\alpha-1)CE+(K-1)(2\beta-1)DF}{K-3}
\end{equation}
then $\frac{\partial x_S}{\partial K}\le 0$, the location $x_S$ of
the domain wall is decreased as the parameter $K(\ge 3)$ increases
(Figure \ref{Figure 13}). Since
\begin{equation}\label{eq34}
\frac{(K^2-1)(2\alpha-1)CE+(K-1)(2\beta-1)DF}{K-3}\ge 0\qquad 0\le
\alpha, \beta\le \frac{1}{K+1}
\end{equation}
the breakdown of the inequality (\ref{eq33}) is difficult to be
found in the computer simulations.

\begin{figure}
  \includegraphics[width=200pt]{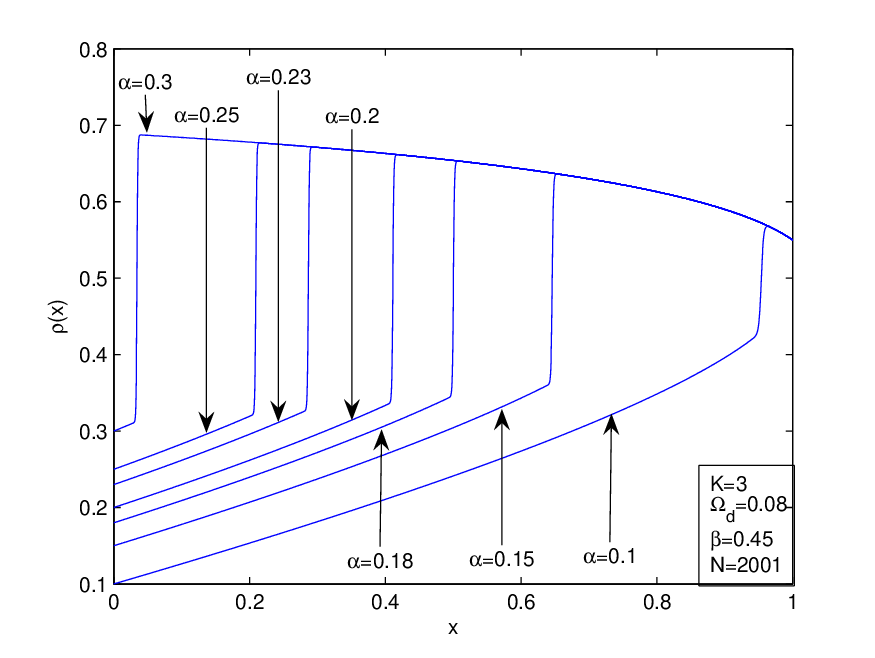}\includegraphics[width=200pt]{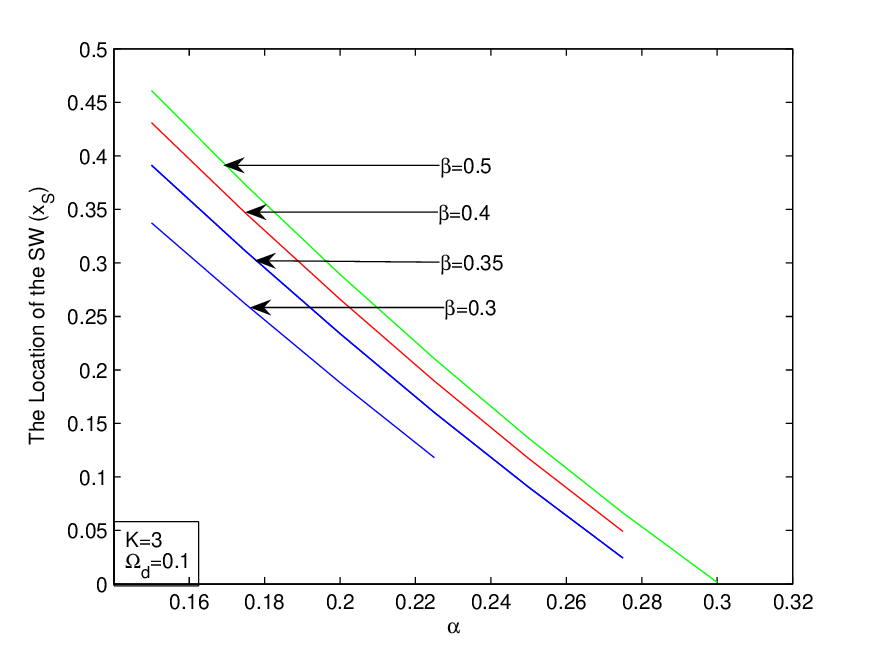}
  \caption{The location $x_S$ of the domain wall is decreased as the increase of
the parameter $\alpha$ for $ \frac{1}{K+1}\le \beta\le 0.5$ {\bf
(left)}. The figure of the function $x_S(\alpha)$ for  $
\frac{1}{K+1}\le \beta\le 0.5$
  {\bf (right)}.} \label{Figure 14}
\end{figure}

\begin{figure}
  \includegraphics[width=200pt]{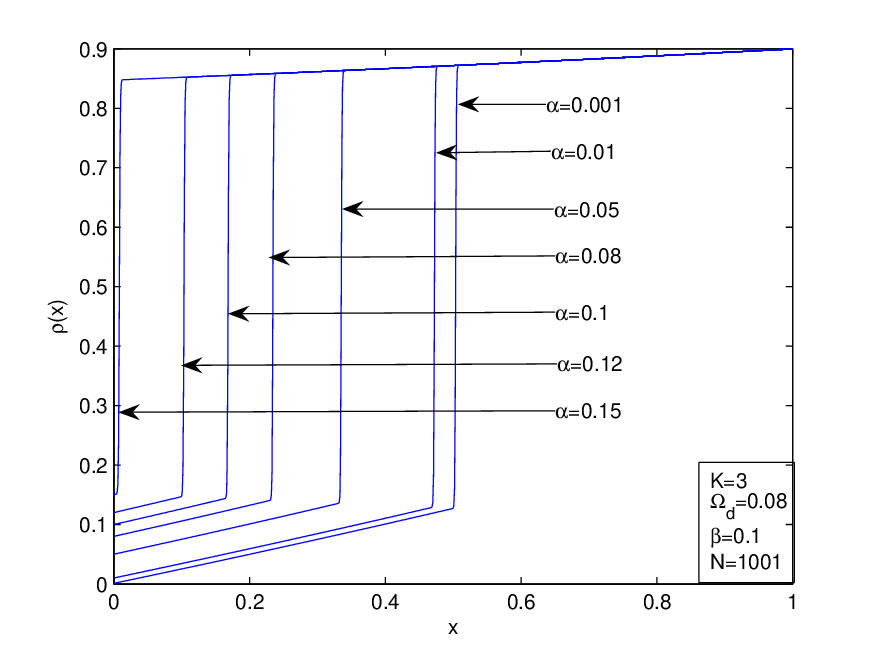}\includegraphics[width=200pt]{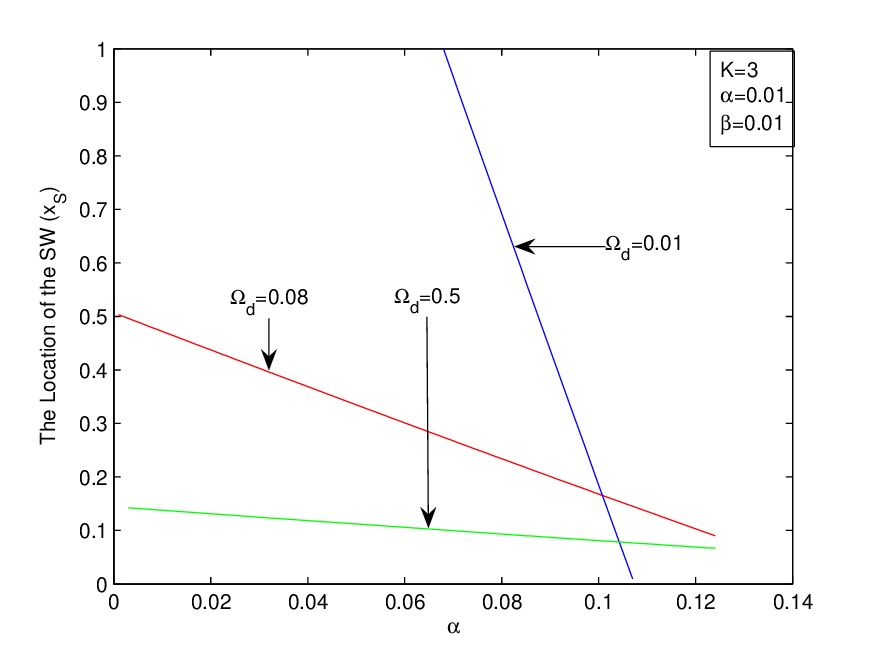}
  \caption{The location $x_S$ of the domain wall is decreased as the increase of
the parameter $\alpha$ for $ 0\le \beta< \frac{1}{K+1}$ {\bf
(left)}.   The figure of the function $x_S(\alpha)$ for  $
 0\le \beta< \frac{1}{K+1}$ {\bf (right)}.} \label{Figure 15}
\end{figure}

\noindent {\bf (c) }
\begin{equation}
\frac{\partial x_S}{\partial
\alpha}=\frac{(1-2\alpha)B}{4\epsilon\Omega_d (K+1)D}\le 0\qquad
\text{for}\ \ \forall 0\le \alpha,\ \beta\le 0.5
\end{equation}
so the location $x_S$ of the domain wall is monotonously decreased
as a function of the parameter $\alpha$ for $ \forall \ K\ge 1, 0\le
\alpha\le 0.5, 0\le \beta\le 0.5$ (the results of the computer
simulations are plotted in Figure \ref{Figure 14}, Figure
\ref{Figure 15})

\begin{figure}
  \includegraphics[width=200pt]{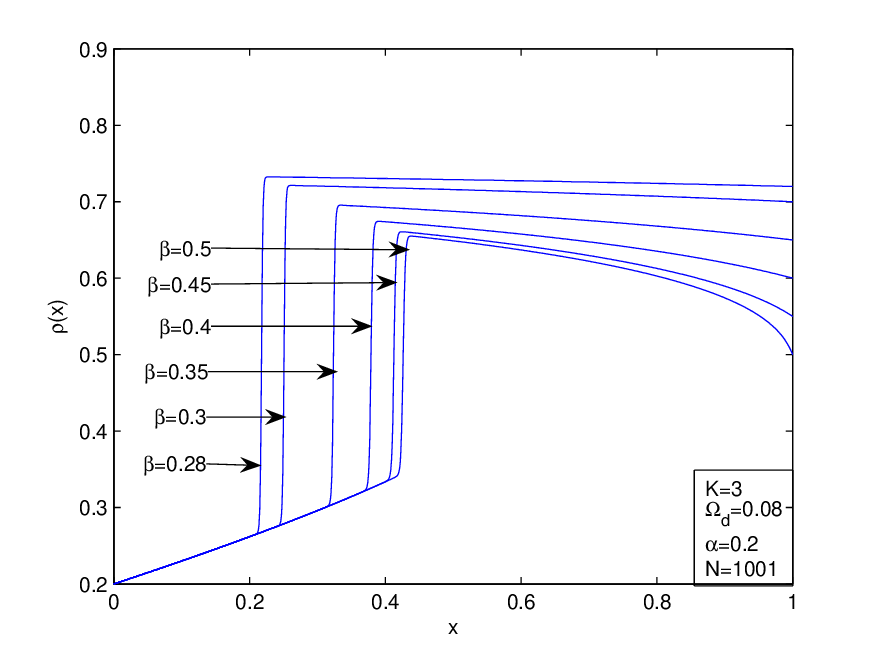}\includegraphics[width=200pt]{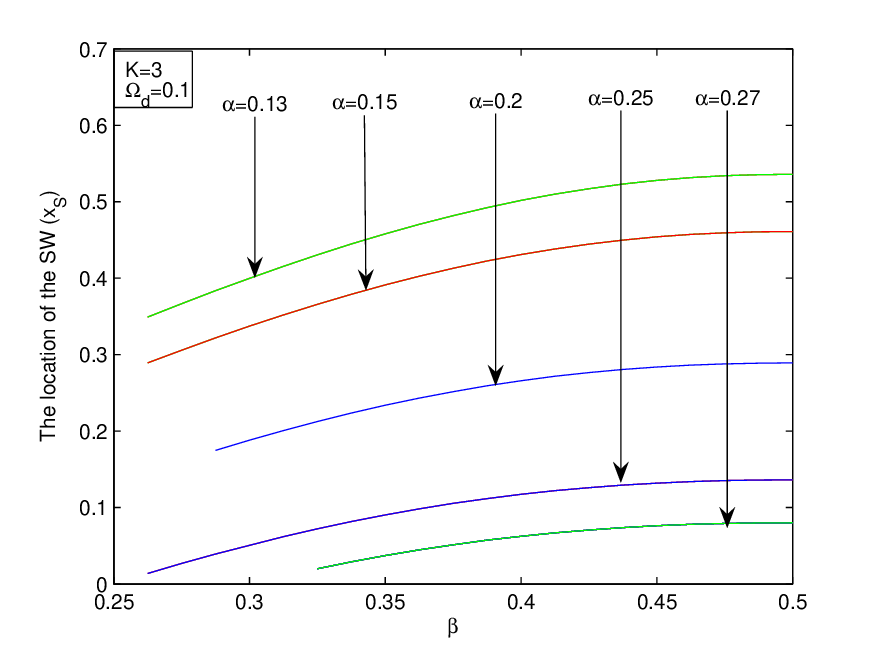}
  \caption{The location $x_S$ of the domain wall is increased as the increase of the parameter $\beta$  for $
\frac{1}{K+1}\le \beta\le 0.5$ {\bf (left)}. The figure of the
function $x_S(\beta)$ for $ \frac{1}{K+1}\le \beta\le 0.5$ {\bf
(right)}.} \label{Figure 16}
\end{figure}

\noindent {\bf (d) }
\begin{equation}
\frac{\partial x_S}{\partial
\beta}=\frac{(2\beta-1)A}{4\epsilon\Omega_d (K+1)C}\ge 0\qquad
\text{for}\ \ \forall 0\le\alpha,\ \beta\le 0.5
\end{equation}
in fact, $A\le 0, C\ge 0$ if $\frac{1}{K+1}\le\beta\le 0.5 $; and
$A\ge 0, C\le 0$ if $\frac{1}{K+1}\le\beta\le 0.5 $. So the location
$x_S$ of the domain wall is monotonously increased as a function of
the parameter $\beta$ for $ \forall \ K\ge 1, 0\le \alpha\le 0.5,
0\le \beta\le 0.5$ ( Figure \ref{Figure 16}, Figure \ref{Figure
17}).
\begin{figure}
  \includegraphics[width=200pt]{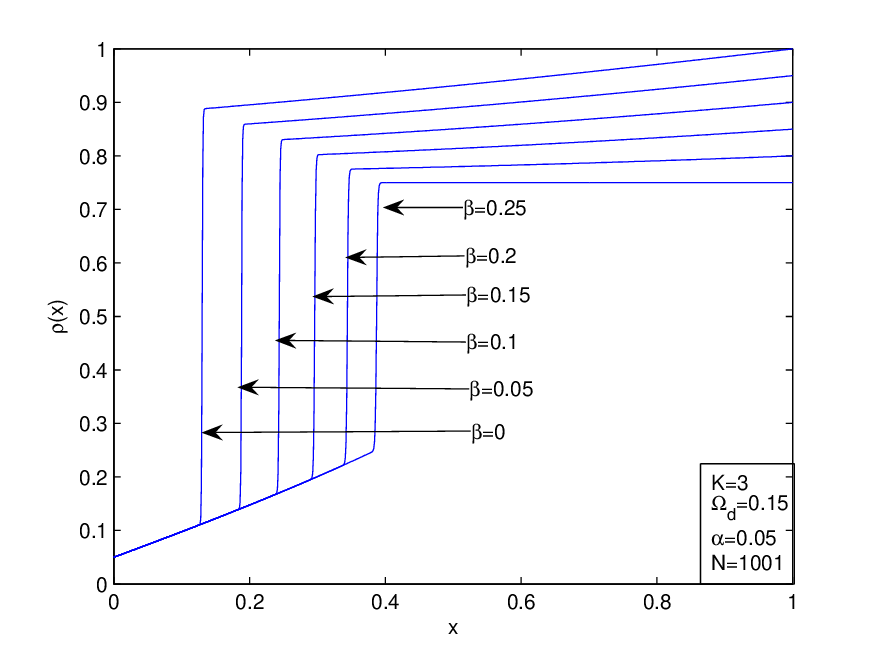}\includegraphics[width=200pt]{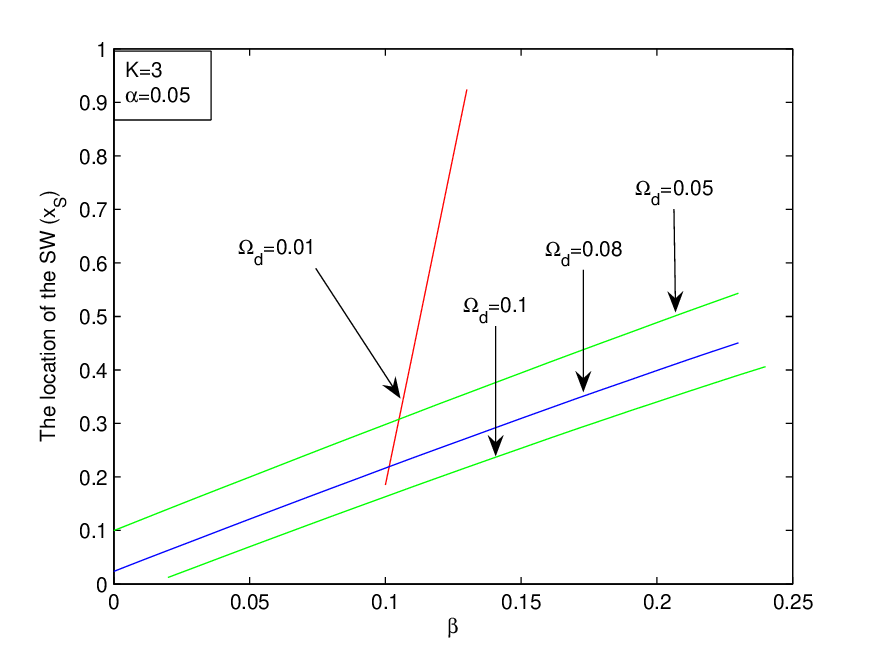}
  \caption{The location $x_S$ of the domain wall is increased as the increase of
the parameter $\beta$ for $ 0\le \beta\le \frac{1}{K+1}$ {\bf
(left)}.  The figure of the function $x_S(\beta)$ for $ 0\le
\beta\le \frac{1}{K+1}$
 {\bf (right)}.} \label{Figure 17}
\end{figure}

\subsection{The Properties of the Height of the domain wall }

\begin{figure}
  \includegraphics[width=200pt]{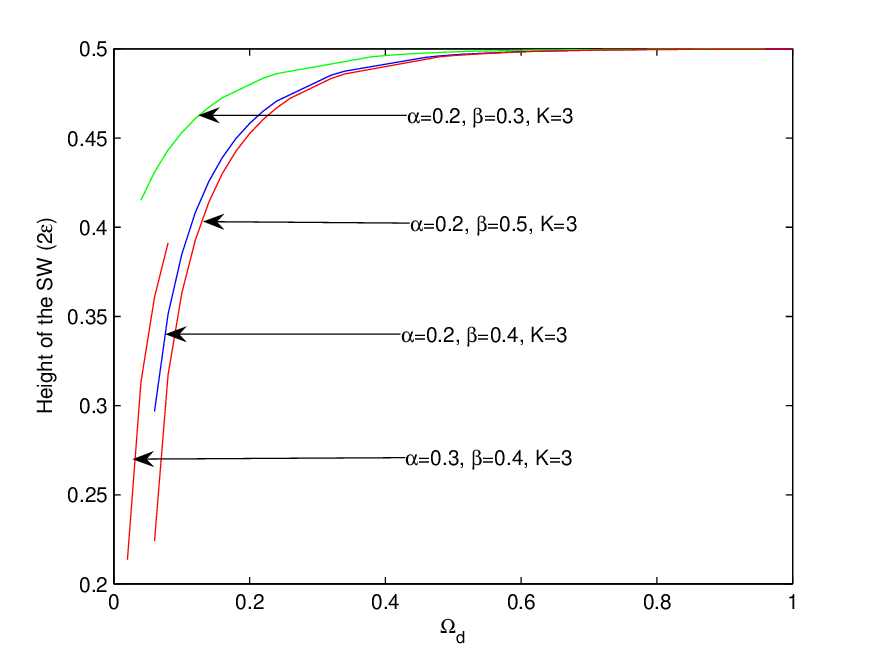}\includegraphics[width=200pt]{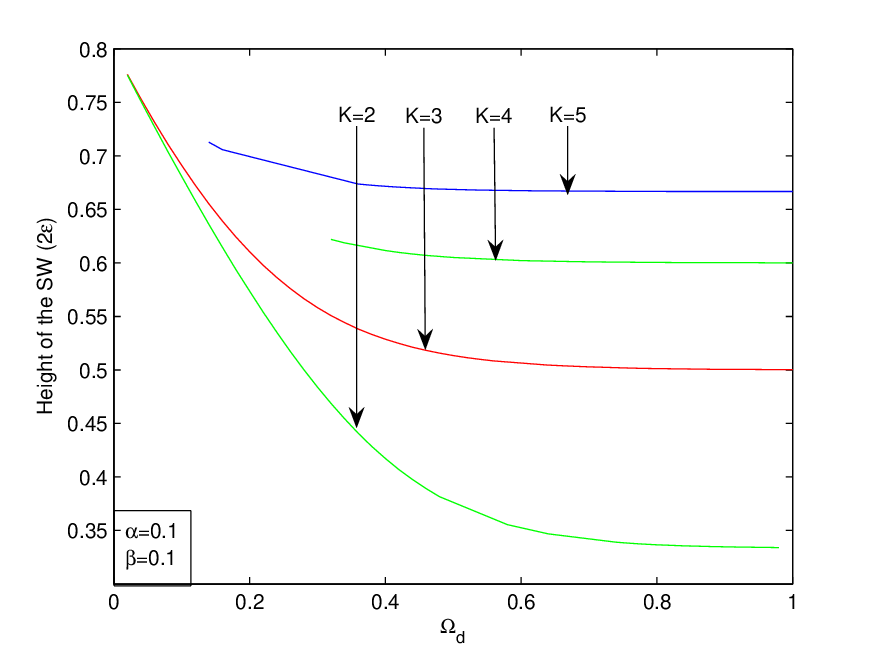}
  \caption{The figure of the function $2\epsilon(\Omega_d)$ for $\frac{1}{K+1}\le \beta\le 0.5$ {\bf (left)}.
 The figure of the function $2\epsilon(\Omega_d)$ for $0\le \beta\le \frac{1}{K+1}${\bf (right)}.} \label{Figure 18}
\end{figure}

In view of the Eqns. (\ref{slol}) (\ref{slor}) and the definition of
the height of the domain wall $2\epsilon$, using the chain rule of
the derivative, we can get the following theoretical results:

\noindent {\bf (a) }
\begin{equation}
\frac{\partial \epsilon}{\partial
\Omega_d}=-\frac{AB}{16(K+1)\epsilon^2}\left\{\begin{aligned} &\ge
0\quad \text{for}\quad &\frac{1}{K+1}\le \beta\le 0.5\cr &\le 0\quad
\text{for}\quad &0\le \beta\le \frac{1}{K+1}\ \
\end{aligned}
\right.
\end{equation}
It can be verified that $A\le 0,\ B\ge 0$ for $\frac{1}{K+1}\le
\beta\le 0.5$; and $A\ge 0,\ B\ge 0$ for $ 0\le \beta\le
\frac{1}{K+1}$. So the height $2\epsilon$ of the domain wall is
monotonously increased as a function of the parameter $\Omega_d$ for
$\frac{1}{K+1}\le \beta\le 0.5$, and monotonously decreased as a
function of the parameter $\Omega_d$ for $ 0\le \beta\le
\frac{1}{K+1}$ (Figure \ref{Figure 9} {\bf (left)}, Figure
\ref{Figure 10} {\bf (left)}, Figure \ref{Figure 11} {\bf (left)},
Figure \ref{Figure 18}).

\begin{figure}
  \includegraphics[width=200pt]{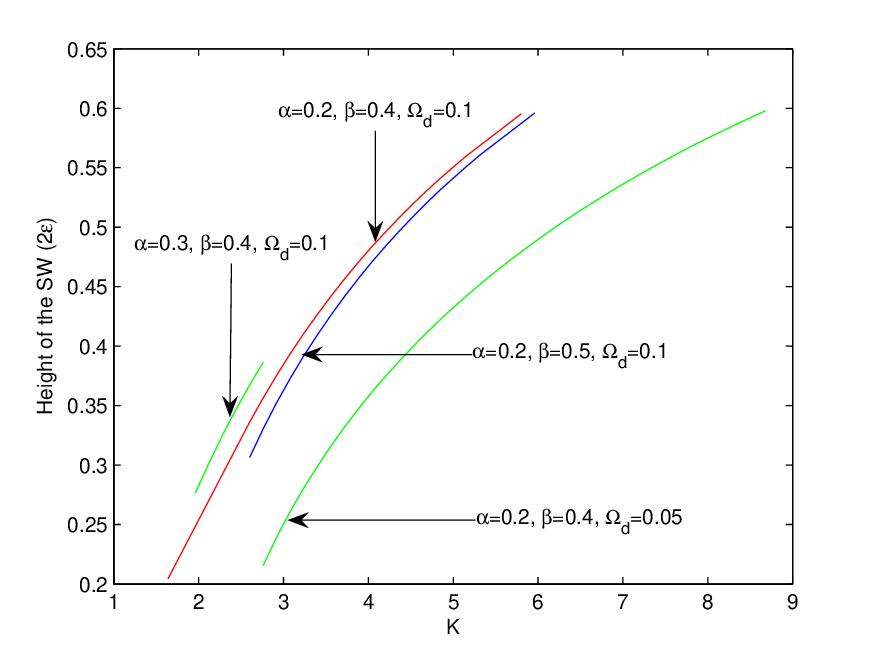}\includegraphics[width=200pt]{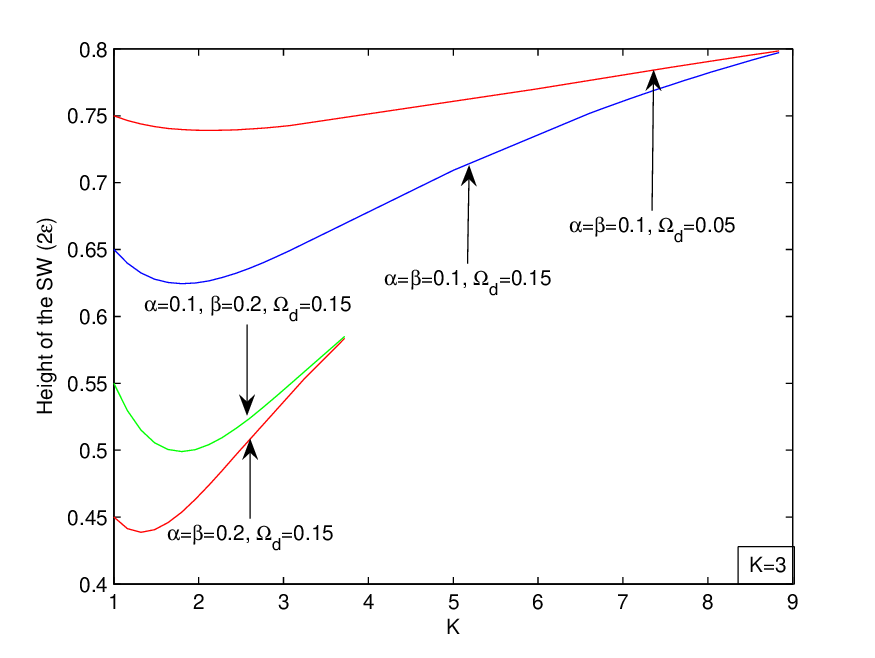}
  \caption{The figure of the function $2\epsilon(K)$ for $\frac{1}{K+1}\le \beta\le 0.5$ {\bf (left)}.
 The height $2\epsilon$ of the domain wall is not a monotone function of the parameter $K$ for $ 1<K\le 3,\ 0\le \beta< \frac{1}{K+1},\ 0\le
\alpha\le 0.5${\bf (right)}.} \label{Figure 19}
\end{figure}

\noindent {\bf (b) }
\begin{equation}
\begin{aligned}
\frac{\partial \epsilon}{\partial
K}=&-\left\{\frac{[(K-1)^2+2BD]ACE+[(K-1)^2+2AC]BDF}{(K+1)^2(K-1)CD}\right.\cr
&+\left.\frac{(K-3)AB\Omega_d}{K^2-1}\right\}\frac{1}{16(K+1)\epsilon^2}
\end{aligned}
\end{equation}
where
\begin{equation}
\begin{aligned}
&(K-1)^2+2BD=(K^2-1)(2\alpha-1)+4\epsilon(K+1)D\cr
&(K-1)^2+2AC=4\epsilon(K+1)C+(K-1)(1-2\beta)
\end{aligned}
\end{equation}
From (15) and (24), one can know
\begin{equation}
\frac{\partial \epsilon}{\partial K}\ge 0\qquad\text{for}\ K\ge 3,\
\frac{1}{K+1}\le \beta\le 0.5,\ 0\le \alpha\le 0.5
\end{equation}
so the height $2\epsilon$ of the domain wall is monotonously
increased as a function of the parameter $K$ for $ K\ge 3,\
\frac{1}{K+1}\le \beta\le 0.5,\ 0\le \alpha\le 0.5$ (Figure
\ref{Figure 12} {\bf (left)}, Figure \ref{Figure 19} {\bf (left)}).
However, for $0\le \beta< \frac{1}{K+1} $, the height $2\epsilon$ of
the domain wall is not a monotone function of parameter $K$ (Figure
\ref{Figure 19} {\bf (right)}).

\begin{figure}
  \includegraphics[width=200pt]{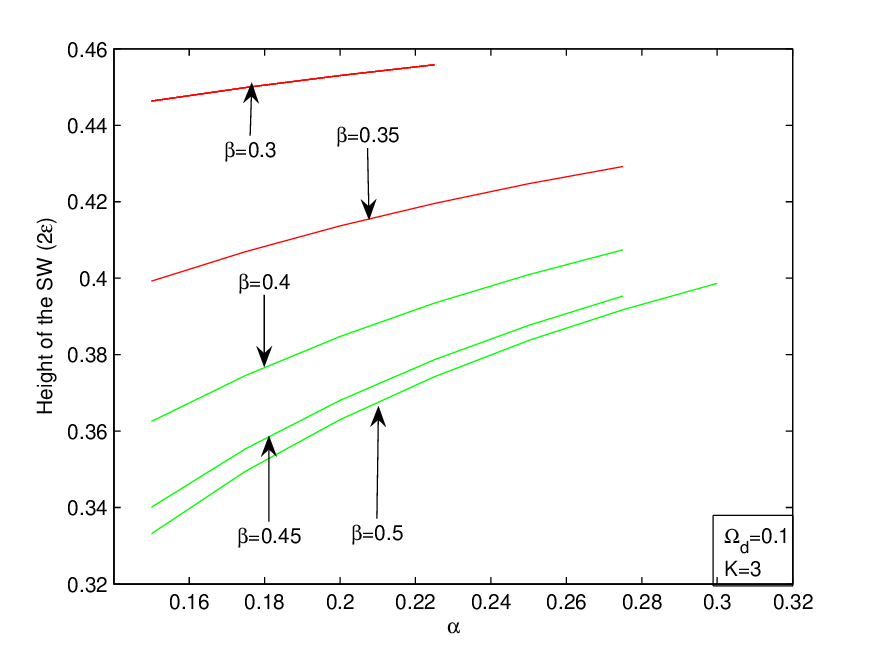}\includegraphics[width=200pt]{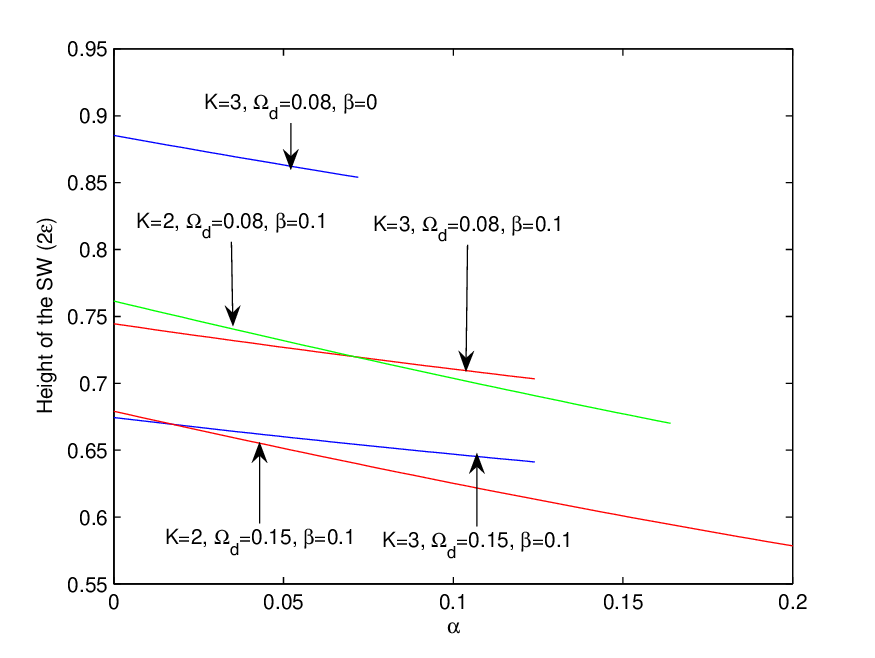}
  \caption{The figure of the function $2\epsilon(\alpha)$ for $\frac{1}{K+1}\le \beta\le 0.5$ {\bf (left)}.
 The figure of the function $2\epsilon(\alpha)$ for $0\le\beta\le\frac{1}{K+1}${\bf (right)}.} \label{Figure 20}
\end{figure}

\noindent {\bf (c) }
\begin{equation}
\frac{\partial \epsilon}{\partial
\alpha}=\frac{(1-2\alpha)AB}{16(K+1)^2\epsilon^2D}\left\{
\begin{aligned}
&\ge 0\qquad &\frac{1}{K+1}\le \beta\le 0.5\cr &\le 0 &0\le \beta\le
\frac{1}{K+1}\ \
\end{aligned}
\right.
\end{equation}
so the height $2\epsilon$ of the domain wall is monotonously
increased as a function of the parameter $\alpha$ for
$\frac{1}{K+1}\le \beta\le 0.5$, and monotonously decreased as a
function of the parameter $\alpha$ for $0\le \beta\le \frac{1}{K+1}$
(Figure \ref{Figure 14}{\bf (left)}, Figure \ref{Figure 15} {\bf
(left)}, Figure \ref{Figure 20}).

\begin{figure}
  \includegraphics[width=200pt]{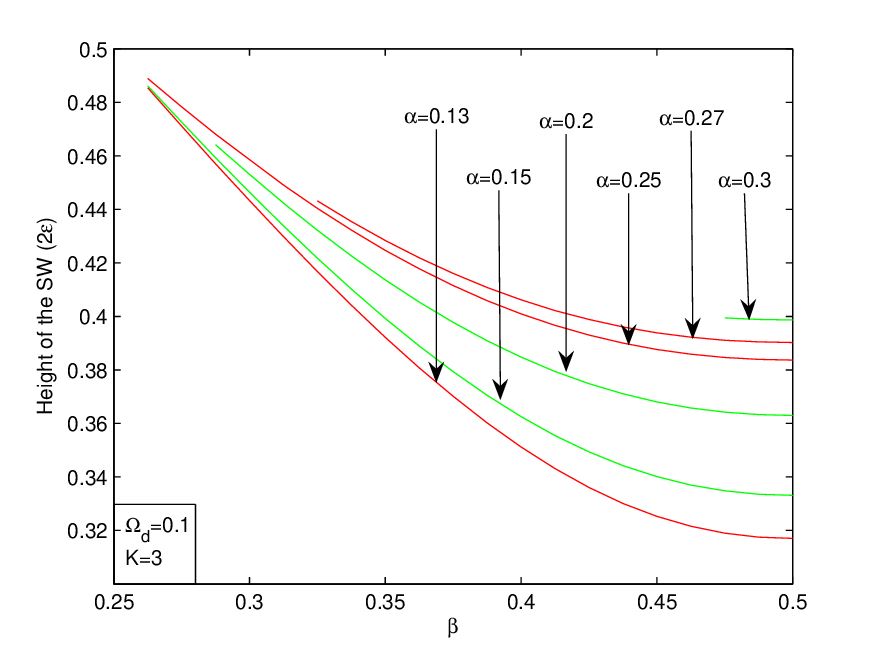}\includegraphics[width=200pt]{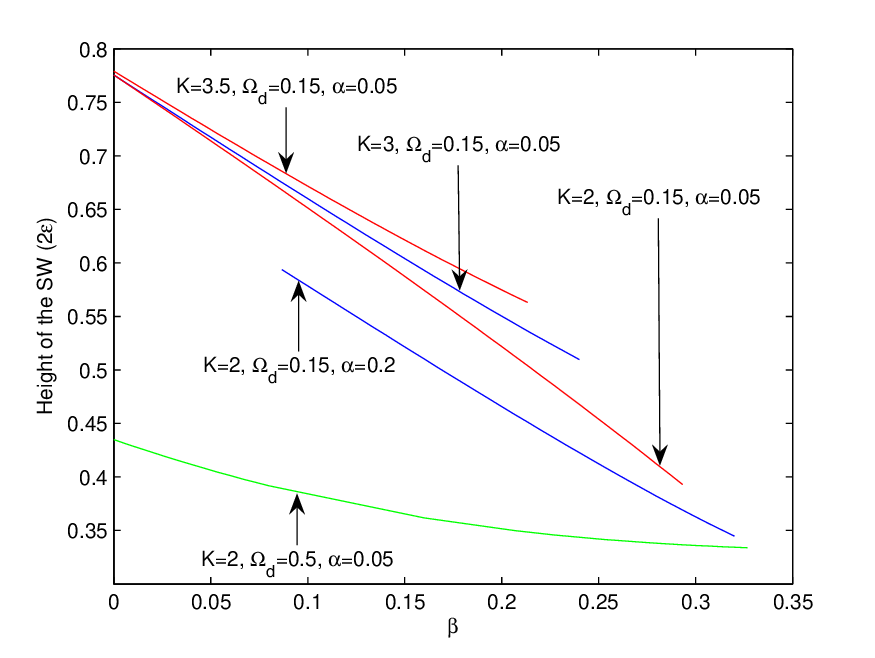}
  \caption{The figure of the function $2\epsilon(\beta)$ for $\frac{1}{K+1}\le \beta\le 0.5$ {\bf (left)}.
 The figure of the function $2\epsilon(\beta)$ for $0\le\beta\le\frac{1}{K+1}${\bf (right)}.} \label{Figure 21}
\end{figure}

\noindent {\bf (d) }
\begin{equation}
\frac{\partial \epsilon}{\partial
\beta}=\frac{(1-2\beta)AB}{16(K+1)\epsilon^2C}\le 0\qquad\text{for}\
0\le \alpha,\ \beta\le 0.5
\end{equation}
since $\frac{A}{C}\le 0$ and $B\ge 0$ for $\forall\ 0\le \alpha,\
\beta\le 0.5$. Therefore, the height $2\epsilon$ of the domain wall
is monotonously decreased as a function of the $\beta$ for $\forall
0\le \beta\le 0.5$ (Figure \ref{Figure 16} {\bf (left)}, Figure
\ref{Figure 17} {\bf (left)}, Figure \ref{Figure 21}).

\section{Conclusions}
In this research, the totally asymmetric exclusion process without
particle number conservation in large particle number limit have
been studied theoretically. Two questions are answered completely:
(1) when and where does the domain wall exist? (2) How do the
location and height of the domain wall change as the parameters
$\alpha, \beta, K, \Omega_d$ change? (see {\bf Table 1}, where
$\uparrow$ ($\downarrow$) means the function is a monotonously
increased (decreased) one, $\uparrow\downarrow$
($\downarrow\uparrow$) means the function has an unique maximum
(minimum) point, $\downarrow ?$ means that the function might be
monotonously decreased, but the proof is not completed in this
research)

In summery, we have found: (1)  For $0\le\alpha\le 0.5, 0\le
1-\beta\le\frac{K}{K+1}$, there exists domain wall if and only if $
\rho^{-1}_{l\alpha}(\gamma)\le 1\ \text{and}\
\rho^{-1}_{r\gamma}(1-\alpha)\le 0,\  \text{where}\
\gamma=\min(0.5,\ \beta) $. (2) For $0\le\alpha\le 0.5,
\frac{K}{K+1}\le 1-\beta\le 1$, there exists domain wall if and only
if $\rho_l^{-1}(\beta)\le 1\quad \text{and}\quad
\rho_r^{-1}(1-\alpha)\ge 0$. (3) For $0.5\le\alpha\le 1$, the domain
wall doesn't exist. (4) The location $x_S$ of the domain wall is
monotonously increased (decreased) as a function of the parameter
$\beta$ ($\alpha$). The height $2\epsilon$ of the domain wall is
monotonously decreased as a function of the parameter $\beta$. (5)
For $\frac{1}{K+1}\le \beta\le 0.5, K\ge 3$, The location $x_S$ of
the domain wall is monotonously decreased as a function of the
parameters $\Omega_d$ and $K$; the height $2\epsilon$ of the domain
wall is monotonously increased as a function of the parameters
$\Omega_d$ and $K$. (6) For $ 0\le \beta\le \frac{1}{K+1}$, , the
height $2\epsilon$ of the domain wall is monotonously decreased as a
function of the parameters $\Omega_d$ and $\alpha$. {\small
  \begin{tabular}{|c|c|c|c|c|}
    \hline
      $0\le\alpha\le 0.5$&$\Omega_d \uparrow$ & $K(\ge 3) \uparrow$ & $\alpha \uparrow$ & $\beta \uparrow $  \\
    \hline
    $\frac{1}{K+1}\le \beta\le 0.5$& $x_S \downarrow,\ 2\epsilon \uparrow$ &  $x_S \downarrow,\ 2\epsilon \uparrow$  &  $x_S \downarrow,\ 2\epsilon \uparrow$  &  $x_S \uparrow,\ 2\epsilon \downarrow$  \\
    \hline
    $ 0\le \beta\le \frac{1}{K+1}$& $x_S\uparrow\downarrow$,\ $2\epsilon \downarrow$ &  $x_S\downarrow?$, $2\epsilon\downarrow\uparrow $  &  $x_S \downarrow,\ 2\epsilon \downarrow$  &  $x_S \uparrow,\ 2\epsilon \downarrow$  \\
    \hline
  \end{tabular}
 \centerline{{\bf Table 1:} The properties of the domain wall.}
} \centerline{}

 Recently, totally asymmetric exclusion processes with
internal states and particle detachment and attachment, which serve
as generic transport models in various context, have been introduced
and extensively studied (\cite{Reichenbach2006} \cite{Greulich2007}
\cite{Nishinari2005}). Using the similar methods as in this
research, these generalized models also can be studied
theoretically, and the corresponding results will be given in the
future.

\vskip 0.5cm

\noindent{\bf Acknowledgments}

This work was funded by National Natural Science Foundation of China
(Grant No. 10701029).



\end{document}